\documentclass[5p,times,twocolumn,number,sort&compress]{elsarticle}
\usepackage{changes}
\usepackage{tabularx}
\usepackage{booktabs}
\usepackage{dcolumn} 
\usepackage{graphicx}
\usepackage{multirow}
\usepackage{adjustbox}
\usepackage[T1]{fontenc}
\usepackage[utf8]{inputenc}
\usepackage{color}
\usepackage{dsfont}
\usepackage{tikz}
\usepackage{filecontents}
\usepackage{pgfplots}
\usepgfplotslibrary{fillbetween}
\usetikzlibrary{spy,shadows}
\usetikzlibrary{backgrounds}
\usetikzlibrary {arrows.meta}
\usepgfplotslibrary{groupplots}
\pgfplotsset{compat=newest}
\usepgfplotslibrary{units}
\usepackage{lipsum}
\usepackage{bm}        
\usepackage{amssymb,amsmath,amsfonts}
\usepackage{mathtools, cuted}
\usepackage[breaklinks=true,debug=true]{hyperref}
\usepackage{braket}    

\journal{Physics Letters B}

\begin{document}

\begin{frontmatter}

\title{Understanding the mirror asymmetry in Gamow-Teller transition rates between $^{28}$Al($\beta^-$)$^{28m}$Si and $^{28}$P($\beta^+$)$^{28m}$Si}

\author[first]{L.\,Xayavong}
\ead{xayavong.latsamy@yonsei.ac.kr}
\author[second]{N.\,A.\,Smirnova}
\ead{smirnova@lp2ib.in2p3.fr}
\author[first]{Y.\,Lim\corref{cor1}}
\ead{ylim@yonsei.ac.kr}
\cortext[cor1]{Corresponding author}
\affiliation[first]{organization={Department of Physics, Yonsei University},
            city={Seoul},
            postcode={03722}, 
            country={South Korea}}
\affiliation[second]{organization={LP2IB (CNRS/IN2P3-Universit\'e de Bordeaux)},
            city={Gradignan cedex},
            postcode={33170}, 
            country={France}}

\begin{abstract}
The long-standing discrepancy between shell-model and experimental values of the mirror asymmetry in Gamow-Teller transition rates for $^{28}$Al($\beta^-$)$^{28m}$Si and $^{28}$P($\beta^+$)$^{28m}$Si is partially resolved by integrating the $p$-orbital contribution into the radial mismatch correction term. This approach offers a potential estimate of the radial mismatch-induced core-orbital contribution based on the French-Macfarlane sum rules, avoiding the need for heavy calculations required by the first-principle shell model. Additionally, the isospin-mixing contribution undergoes considerable improvement through scaling the calculated isospin-mixing correction with the experimental energy separation between the lowest two admixed states, in accordance with the result from the two-level model. 
\end{abstract}



\end{frontmatter}


\section{Introduction}\label{introduction}

Nuclear $\beta$ decay between bound states offers a precise means for studying properties of fundamental interactions. This weak semileptonic process in nuclei is typically categorized into Fermi and Gamow-Teller transitions, primarily governed by the weak vector and axial-vector currents, respectively. The coupling constant for the vector current can be uniquely extracted from a pure Fermi transition, such as the superallowed $0^+\rightarrow0^+$ nuclear $\beta$ decay\,\cite{XaNa2022,XaNa2018}, as it is not renormalized in the nuclear medium, due to the conservation of the vector current, known as the CVC hypothesis. 
On the other hand, the coupling constant for the axial-vector current cannot be precisely extracted from Gamow-Teller (GT) transitions, because the GT matrix elements strongly depend on structural details of the initial- and final-state wave functions. 

Nevertheless, although a single GT transition does not seems informative for a precision study of fundamental interactions, the mirror asymmetry in GT transition rates could be attributed to the existence of weak second-class currents, absent in the standard model. This phenomenon has been extensively studied in the past, typically before the year 2023\,\cite{Smirnova2003,TOWNER1973589,WILKINSON1972289,BARKER1992147,PhysRevLett.27.1018}, while contemporary approaches now rely on correlation measurements that are entirely independent on nuclear structure\,\cite{GONZALEZALONSO2019165}. Given that the beyond stand-model contribution is negligibly small compared to nuclear structure-calculation uncertainty\,\cite{PhysRevLett.128.202502,xayavong2023shellmodel}, the asymmetry parameter can be defined as 
\begin{equation}\label{delta}
    \delta=\frac{ft^+}{ft^-}-1\approx (\delta_C^+-\delta_C^-)-(\delta_{NS}^+-\delta_{NS}^-)-(\delta_R'^+-\delta_R'^-), 
\end{equation}
where the label $+$\,($-$) indicate the $\beta^+$\,($\beta^-$) emission. $ft$ is the product of the statistical rate function $(f)$ and the partial half-life $(t)$. The rest includes $\delta_C$, representing the correction due to isospin-symmetry breaking, while $\delta_{NS}$ and $\delta_R'$ account for nuclear structure-dependent and nucleus-dependent radiative effects, respectively. It should be noted that higher-order terms in the corrections are neglected in the derivation of Eq.\,\eqref{delta}. Therefore, Eq.\,\eqref{delta} can serve as a collective test for all the theoretical corrections listed above, using experimental data on $ft^+/ft^-$. 

Within the shell-model framework, $\delta_C$ can be decomposed into two components:  
\begin{equation}
    \delta_C\approx\delta_{C1} + \delta_{C2}, 
\end{equation}
where $\delta_{C1}$ accounts for configuration admixtures within the shell-model valence space induced by the isospin-nonconserving part of the effective Hamiltonian. Whereas, $\delta_{C2}$ accounts for the mismatch between proton and neutron radial wave functions, compensating for the isospin admixtures extending beyond the shell-model valence space. The interference terms, discussed in Ref.\,\cite{xayavong2023shellmodel}, are of higher orders and will be neglected in the present study. 

The isospin-symmetry breaking contribution was recently reevaluated within the shell-model approach for various pairs of mirror GT transitions from $p$ to $sd$ shells\,\cite{xayavong2023shellmodel}. 
Despite remarkable improvements have been made, there are still a number of cases where the deviations from experimental data are several standard deviations. 
Notably, the mirror asymmetry between the GT transition rates of $^{28}$Al($\beta^-$)$^{28m}$Si and $^{28}$P($\beta^+$)$^{28m}$Si was measured to be -3.706$\pm$0.444\,\%\,\cite{NNDCENSDF,xayavong2023shellmodel}, while the calculation using the existing well-established phenomenological effective interactions\,\cite{USD,USDab,OrBr1989x} in the $sd$ model space produces 7.716$\pm$2.961\,\%. We notice that, in the absence of the radiative contribution, a negative asymmetry would indicate a larger isospin-symmetry breaking correction in the $\beta^-$ decay compared to its $\beta^+$ partner, as evident from Eq.\,\eqref{delta}. This phenomenon is rather uncommon as the isospin-symmetry breaking effect is expected to be stronger in proton-rich nuclei, given that the Coulomb repulsion strength increases with the atomic number. Moreover, $\delta_{C2}$ for proton-rich nuclei can be enhanced by the weakly-bound effect. 

The aim of this article is twofold. Firstly, to improve the isospin-mixing correction term for the mirror GT transitions mentioned in the previous paragraph, by scaling calculated values with experimental energy separation between the two lowest admixed states. This empirical method is based on the argument coming out from a two-level model analysis and the first-order perturbation theory\,\cite{xayavong2023shellmodel}. Secondly, to investigate the $p$-orbital contribution by using the French-Macfarlane sum rules\,\cite{FRENCH1961168} on spectroscopic factors and various assumptions for testing the distribution of spectroscopic factors as a function of excitation energy for deep-lying orbitals. 

\section{Empirical refinement for $\delta_{C1}$}\label{sec2}

A detailed study of the isospin-mixing correction term for GT transitions within the shell-model framework was recently carried out in Ref.\,\cite{xayavong2023shellmodel}. 
The general shell-model expression for $\delta_{C1}$ is given by 
\begin{equation}\label{C1x}
\delta_{C1} = \frac{2}{\mathcal{M}_0} \sum_{k_ak_b} \theta_{ab}^{\lambda}\xi_{ab} D(abfi\lambda),
\end{equation}
where the labels $f$ and $i$ denote the final and initial many-particle states, respectively, and $k_{a/b}$ stands for the set of spherical quantum numbers $(nlj)$ of the respective adjacent single-particle states $a$ and $b$. $\theta_{ab}^{\lambda}$ and $\xi_{ab}$ represent the spin-angular and isospin components of the single-particle matrix elements of the $\bm{\sigma}\bm{\tau}_{\pm}$ operator, respectively. Their complete expressions can be found in Ref.\,\cite{xayavong2023shellmodel}. $D(abfi\lambda)$ denotes the deviation of the one-body transition density from the corresponding isospin-symmetry value. $\mathcal{M}_0$ is the isospin-symmetry GT matrix element. Unlike Fermi transitions, $\mathcal{M}_0$ for an axial-vector process cannot be obtained analytically. For the considered mirror GT transitions, the averaged $\mathcal{M}_0$ value for the three effective universal $sd$-shell interactions\,\cite{USD,USDab} is 0.751$\pm$0.058. 

\begin{center}
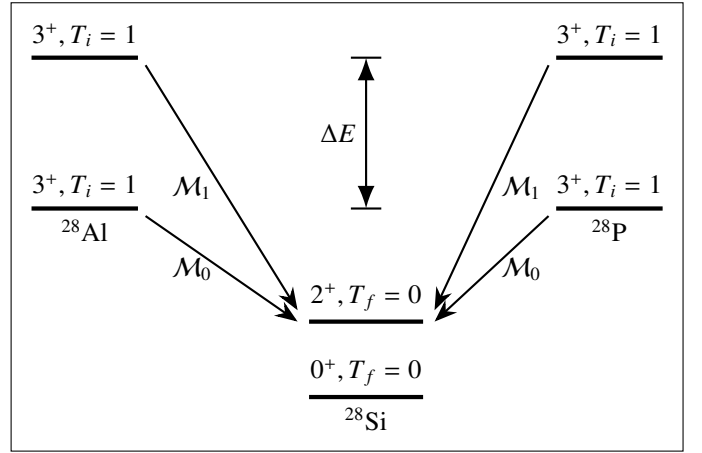
\begin{figure}[ht!]
  \begin{tikzpicture}[framed]
  \draw[ultra thick] (-4.4,3.5) -- (-3,3.5) node[above,midway] {$3^+,T_i=1$} ; 
  \draw[ultra thick] (-4.4,1.5) -- (-3,1.5) node[above,midway] {$3^+,T_i=1$} node[below,midway] {$^{28}$Al};
  \draw[ultra thick] (3.9,3.5) -- (2.5,3.5) node[above,midway] {$3^+,T_i=1$} ; 
  \draw[ultra thick] (3.9,1.5) -- (2.5,1.5) node[above,midway] {$3^+,T_i=1$} node[below,midway] {$^{28}$P};
  \draw[thick,{Latex[length=3mm]}-{Latex[length=3mm]}] (0,1.5) -- (0,3.5) node[midway,left] {$\Delta E$};
  \draw[thick] (-0.2,1.5) -- (0.2,1.5); 
  \draw[thick] (-0.2,3.5) -- (0.2,3.5); 
  \draw[ultra thick] (-0.75,0) -- (0.75,0) node[above,midway] {$2^+,T_f=0$} ; 
  \draw[ultra thick] (-0.75,-1) -- (0.75,-1) node[above,midway] {$0^+,T_f=0$} node[below,midway] {$^{28}$Si};
  \draw[thick,-{Stealth[length=3mm]}] (-2.9,1.4) -- (-0.9,0) node[midway,left] {$\mathcal{M}_0$}; 
  \draw[thick,-{Stealth[length=3mm]}] (-2.9,3.4) -- (-0.9,0.15) node[midway,left] {$\mathcal{M}_1$};
  \draw[thick,-{Stealth[length=3mm]}] (2.4,3.4) -- (0.9,0.15) node[right,midway] {$\mathcal{M}_1$}; 
  \draw[thick,-{Stealth[length=3mm]}] (2.4,1.4) -- (0.9,0) node[right,midway] {$\mathcal{M}_0$};
  \end{tikzpicture}
\caption{Schematic representation of the mirror pair of Gamow-Teller transitions considered in the present study.}
\label{fig1}
\end{figure}
\end{center}

In this shell-model approach, all structural information of the initial- and final-state wave functions is encapsulated in $D(abfi\lambda)$. This quantity often becomes intricate, because the shell-model configuration mixing could occur across a vast number of states, depending on the size of valence space, the number of active particles and the nature of the driven interactions. On the other hand, even though performing a full model space diagonalization of the Hamiltonian matrix is feasible, an exact shell-model calculation of $\delta_{C1}$ strongly depends on the effective isospin-nonconserving interaction. In general, the existing interactions produced through a global fit to binding and excitation energies, and the coefficients of the isobaric multiplet mass equations, are not able to provide a consistently precise description for all nuclei resided in the corresponding model space, in particular those near the drip-lines. In order to gain deeper insight into this correction term, one can employ a two-level model as described in full details in Ref.\,\cite{xayavong2023shellmodel}. Within this simplified model, an analytic formula for $\delta_{C1}$ is derived to be, 
\begin{equation}\label{C1}
\displaystyle \delta_{C1} = -2\eta\alpha + \left( 1-\eta^2 \right) \alpha^2 + \mathcal{O}(\alpha^3), 
\end{equation}
where $\eta=\mathcal{M}_1/\mathcal{M}_0$, representing the ratio of the isospin-symmetry GT matrix elements between the upper (labeled as 1) and the lower (labeled as 0) admixed states. 
Evidently, the presence of $\eta$ in Eq.\,\eqref{C1} differentiates GT from Fermi transitions, as $\eta=0$ for a pure vector process due to the isospin forbiddance of Fermi transitions to nonanalog states. In addition, the presence of $\eta$ intensifies sensitivity to nuclear structure. For instance, the GT matrix elements $\mathcal{M}_0$ and $\mathcal{M}_1$ cannot be obtained analytically, although they solely determined by the isospin-symmetry component of the nuclear Hamiltonian. 
As a result, calculating $\delta_{C1}$ for GT transitions is generally more challenging compared to Fermi transitions.

According to Ref.\,\cite{xayavong2023shellmodel}, the averaged $\delta_{C1}$ values are 0.506$\pm$0.087\,\% for $^{28}$Al($\beta^-$)$^{28}$Si and 0.835$\pm$0.309\,\% for $^{28}$P($\beta^+$)$^{28}$Si. Their contribution to the asymmetry $\delta$ is 0.329$\pm$0.279\,\%. Although these numbers seem to be small, the uncertainty is almost 85\,\% of the mean value. 
In these shell-model calculations, the isospin-nonconserving component for all the aforementioned effective interactions is that produced in Ref.\,\cite{OrBr1989x}. Therefore, this uncertainty reflects the spread due to the difference in the isospin-symmetry components of the interactions. 

Given that, at least, the predicted order of $\delta_{C1}$ by the shell model is correct, the higher-order term in $\alpha$ in Eq.\,\eqref{C1} would be insignificant. Furthermore, according to the first-order perturbation theory, the mixing amplitude $\alpha$ is known to be
\begin{equation}
\alpha=\frac{\braket{\Phi_1|V_{INC}|\Phi_0}}{\Delta E}
\end{equation}
where $V_{INC}$ is the isospin-nonconserving part of the nuclear Hamiltonian, and $\Delta E$ the energy separation between the admixed states. Given that $\Delta E$ is not significantly affected by the isospin mixing, it can be replaced with the experimental data\,\cite{NNDCENSDF}. From these results, the calculated $\delta_{C1}$ values can be improved by scaling with the experimental $\Delta E$ as
\begin{equation}
    \delta_{C1}^{scale} = \delta_{C1}^{cal} \times \frac{\Delta E^{cal}}{\Delta E^{exp}}, 
\end{equation} 
where $\delta_{C1}^{cal}$ and $\Delta E^{cal}$ represent the calculated isospin-mixing correction and energy separation, respectively. 

For the present case, the mirror nuclei $^{28}$Al and $^{28}$P decay from their $3^+$ ground states to a common final state, which is the first $2^+$ state in $^{28}$Si. In this situation, the observed mirror asymmetry is only attributed to the difference in the isospin admixture between the $3^+$ ground states of $^{28}$Al and $^{28}$P. The schematic representation of this process is illustrated in Fig.\,\ref{fig1}. Although Eq.\,\eqref{C1} was originally derived with the assumption that isospin mixing occurs only in the final state\,\cite{xayavong2023shellmodel}, it applies equally to the reverse situation, since GT matrix elements do not depend on the decay direction; on the other word, they are invariant under the interchange of the initial with final states. The isospin-symmetry parameters $\mathcal{M}_1$, $\eta$, and $\Delta E^{cal}$ calculated using the three universal $sd$-shell interactions are 0.329$\pm$0.072, 0.441$\pm$0.112, and 0.961$\pm$0.111\,MeV, respectively. The experimental $\Delta E^{exp}$ is taken as the average of the data for $^{28}$Al and $^{28}$P\,\cite{Au2009}. 
The difference between these data is considered as an experimental uncertainty. As a result, the scaled values of $\delta_{C1}$ become 0.453$\pm$0.100\,\% and 0.747$\pm$0.296\,\% for $^{28}$Al($\beta^-$)$^{28}$Si and $^{28}$P($\beta^+$)$^{28}$Si, respectively. These correspond to a mirror-asymmetry contribution of 0.294$\pm$0.312\,\%. Unfortunately, this outcome shows only marginal improvement, as the total theoretical asymmetry still significantly exceeds the experimental value. 

\section{Shell-model calculations of $\delta_{C2}$}\label{sec3}

The radial mismatch correction term $\delta_{C2}$ arises due to the difference between proton and neutron radial wave functions. It indirectly accounts for the isospin-symmetry breaking contribution beyond the shell-model valence space. This correction term is absent in the conventional shell model which employs the isospin-invariant harmonic-oscillator basis. 
A suitable formalism for $\delta_{C2}$ for GT transitions has been developed in Ref.\,\cite{xayavong2023shellmodel}. 
The isospin-formalism $\delta_{C2}$ expression when the initial and final states have different isospin quantum numbers, corresponding to the present case, can be written as, 
\begin{equation}\label{C2}
\begin{array}{ll}
\delta_{C2} &= \displaystyle \frac{(2T+1)}{(T+1)\mathcal{M}_0} \sum_{k_ak_b\tilde{\pi}}\theta_{ab}^\lambda \Theta_{abfi}^{\tilde{\pi}\lambda} \Lambda_{ab}^{\tilde{\pi}}\xi_{ab} \\[0.18in] 
& \displaystyle \times \sqrt{S^T(f;\tilde{\pi} k_a)S^T(i;\tilde{\pi} k_b)},  
\end{array}
\end{equation}
where $T=\min{(T_i,T_f)}$ and $\tilde{\pi}$ denotes intermediate states with $T_\pi = T+\frac{1}{2}$. For the present case, we have $T_i=1$ and $T_f=0$, then $T=0$ and intermediate states with $T_\pi=\frac{1}{2}$ contribute only. Similar to $\theta_{ab}^\lambda $, the factor $\Theta_{abfi}^{\tilde{\pi}\lambda}$ depends solely on angular momentum quantum numbers, defining the GT selection rules. Their complete expressions can be found in Ref.\,\cite{xayavong2023shellmodel}. 
$S^T(f;\tilde{\pi} k_a)$ and $S^T(i;\tilde{\pi} k_b)$ represent the isospin-invariant spectroscopic factors defined in Ref.\,\cite{xayavong2023shellmodel}. $\Lambda_{ab}^{\tilde{\pi}}$ is the radial mismatch factor representing the deviation from unity of the overlap integral between proton and neutron radial wave functions. In an usual approach, the evaluation of $\Lambda_{ab}^{\tilde{\pi}}$ employs Woods-Saxon radial wave functions\,\cite{XaNa2018,xayavong2023shellmodel,PhysRevC.108.064310}. Essentially, the potential depth and length parameter are readjusted such that the experimental data on separation energies and charge radii are reproduced. Note that the label $\tilde{\pi}$ is added to the radial mismatch factor, due to the incorporation of excitation energy of intermediate state $\ket{\tilde{\pi}}$ in this optimization process. For the transitions under consideration, the charge radii of the nuclear states are not available. Therefore, the length parameter is fixed at the standard value\,\cite{XaNa2018} for all the four nuclei. Fortunately, upon our sensitivity study for these GT transitions, $\delta_{C2}$ is not very sensitive to the length parameter. We found that the calculated asymmetry is still inconsistent with the experimental value even if the length parameter is allowed to vary within the range between 1.2\,fm and 1.3\,fm. The Coulomb term in the potential is evaluated with the approximation of a uniformly distributed charged sphere\,\cite{PhysRevC.108.064310}. 

In Ref.\,\cite{xayavong2023shellmodel}, substantial progress was achieved in calculating $\delta_{C2}$ for these transitions. This was accomplished by extending the inclusion of intermediate states to ensure convergence and integrating an extra surface-peaked term within the Woods-Saxon potential. The resulting $\delta_{C2}$ values are 0.631$\pm$0.892\,\% and 8.299$\pm$2.809\,\% for $^{28}$Al($\beta^-$)$^{28}$Si and $^{28}$P($\beta^+$)$^{28}$Si, respectively. The corresponding contribution to the asymmetry parameter is 7.668$\pm$2.947\,\%. 
However, within these results, the overall asymmetry is still considerably larger than the measured value. 

\section{Core-orbital contribution}\label{sec4}

We may interpret that the adjustment of potential parameters for each nucleus serves as an empirical approach to ensure the effectivity of the GT operator in a given valence space. Although, a similar approach works very well for superallowed $0^+\rightarrow0^+$ Fermi transitions\,\cite{XaNa2022,XaNa2018,xayavong2022higherorder,HaTo2020}, its reliability might be deteriorated for GT transitions because of greater sensitivity to details of initial- and final-state structures. We recall that, in the case of superallowed $0^+\rightarrow0^+$ Fermi transitions, the core-orbital contribution to $\delta_{C2}$ can be expected to be negligible, due to the cancellation between the greater- and lesser-isospin components\,\cite{ToHa2008}. However, Eq.\,\eqref{C2} for GT transitions consists only of a single-isospin component. Because of this property, the core-orbital contribution for GT transitions with $T_i\ne T_f$ could produce a substantial impact, even if the corresponding radial mismatch factor itself is insignificant. Note that only the core-orbital contribution induced by the mismatch between proton and neutron radial wave functions is considered throughout this paper. This is deemed reasonable, given that $\delta_{C2}$ is expected to dominate in axial-vector processes. 

Now, we return our focus to the contribution of deeply-bound orbitals for the present case. 
According to the French-Macfarlane sum rules\,\cite{FRENCH1961168}, the sum of the spectroscopic factors for a pick-up reaction can be related to occupation numbers in the target nucleus. Considering the reaction involving the removal of one proton from the ground state of $^{28}$P, the resulting $^{27}$Si states can possess an isospin of either $T_{\pi}=\frac{1}{2}$ or $\frac{3}{2}$. These final states with lesser isospin correspond the intermediate states specified in Eq.\,\eqref{C2} (denoted with $\tilde{\pi}$), while those with greater isospin are not relevant for the present discussion. The sum rule for the lesser isospin can be expressed as 
\begin{equation}\label{FM1}
\sum_{\tilde{\pi}} S^T(i;\tilde{\pi} k_b)=n_{k_b}^p - \frac{n_{k_b}^n}{3}, 
\end{equation}
where $n_{k_b}^p$ and $n_{k_b}^n$ denote the proton- and neutron-occupation numbers of orbital $k_b$ in the initial state, respectively. Note that the label $T$ is added to the spectroscopic factors to indicate that they are evaluated with an isospin-conserving interaction. Similarly, if we consider the reaction for picking up one neutron from the first $2^+$ state of the self-conjugate $^{28}$Si nucleus, the final $^{27}$Si states can only have isospin $\frac{1}{2}$. The corresponding sum rule can be written as 
\begin{equation}\label{FM2}
\sum_{\tilde{\pi}} S^T(f;\tilde{\pi} k_a)=3 n_{k_a}^n, 
\end{equation}
where $n_{k_a}^n$ denotes the neutron-occupation number of orbital $k_a$ in the first $2^+$ state of $^{28}$Si. 
Note that the isospin quantum numbers and the Clebsch-Gordan coefficients have been replaced with their respective numerical values. As a convention, the summations in Eq.\,\eqref{FM1} and \eqref{FM2} progress from low to high energy states. In general, the core orbitals are highly filled. In this study, we approximate their occupation numbers as $n_{k_b}^p\approx n_{k_b}^n\approx 2j_b+1$ and $n_{k_a}^n\approx 2j_a+1$, assuming they are fully filled. We notice that Eq.\,\eqref{FM1} and \eqref{FM2} also apply to the nuclei associated with the mirror $\beta^-$ partner, with the interchange of protons and neutrons, since the spectroscopic factors are mirror invariant. 

It should be noted that, while the energy gap separating the valence space and the inert core may not fully prevent significant cross-shell excitation, the radial mismatch factor for core orbitals is expected to be considerably smaller than for valence orbitals, due to the so called binding-energy effect\,\cite{PhysRevLett.27.1018}. 
Additionally, an intermediate-nucleus excitation increases the separation energies of both the initial and final nuclei. 
As a consequence, the decaying and resulting nucleons become more tightly bound to the mean-field potential generated by the intermediate system, leading to a further reduction in the radial mismatch factor. Furthermore, the spectroscopic factors generally decrease on average with intermediate-state energy. According to these properties, it would be sufficient to limit the sum over $\tilde{\pi}$ for core orbitals to include only a few lowest intermediate states. The $p$-orbital contribution to $\delta_{C2}$ can be decomposed as 
\begin{strip}
\begin{equation}\label{p}
\begin{array}{ll}
\delta_{C2}^{p-shell}=&\displaystyle \sqrt{\frac{35}{3}} \frac{2}{\mathcal{M}_0}\left\{ -\frac{2}{\sqrt{55}}\sum_{\tilde{\pi}}^{J_{\tilde{\pi}}^\pi=\frac{5}{2}^-}\left[\Lambda^{\tilde{\pi}}_{p_\frac{1}{2} p_\frac{3}{2}}\right] - 2\sqrt{\frac{2}{175}}\sum_{\tilde{\pi}}^{J_{\tilde{\pi}}^\pi=\frac{5}{2}^-}\left[\Lambda^{\tilde{\pi}}_{p_\frac{3}{2} p_\frac{1}{2}}\right] -\frac{1}{\sqrt{110}}\sum_{\tilde{\pi}}^{J_{\tilde{\pi}}^\pi=\frac{5}{2}^-}\left[\Lambda^{\tilde{\pi}}_{p_\frac{1}{2} p_\frac{1}{2}}\right] \right. \\[0.2in]
&\left. + \displaystyle \frac{1}{3\sqrt{29}} \sum_{\tilde{\pi}}^{J_{\tilde{\pi}}^\pi=\frac{3}{2}^-}\left[\Lambda^{\tilde{\pi}}_{p_\frac{3}{2} p_\frac{3}{2}}\right] - \frac{2}{3}\sqrt{\frac{35}{715}}\sum_{\tilde{\pi}}^{J_{\tilde{\pi}}^\pi=\frac{5}{2}^-}\left[\Lambda^{\tilde{\pi}}_{p_\frac{3}{2} p_\frac{3}{2}}\right] + \frac{1}{3}\sqrt{\frac{170}{1001}}\sum_{\tilde{\pi}}^{J_{\tilde{\pi}}^\pi=\frac{7}{2}^-}\left[\Lambda^{\tilde{\pi}}_{p_\frac{3}{2} p_\frac{3}{2}}\right] \right\}, 
\end{array}
\end{equation}
\end{strip} 
where the symbol $\left[\Lambda^{\tilde{\pi}}_{k_ak_b}\right]$ is the shorthand for $\Lambda^{\tilde{\pi}}_{k_ak_b}\times\left[S^T(f;\tilde{\pi} k_a)S^T(i;\tilde{\pi} k_b)\right]^\frac{1}{2}$. $J_{\tilde{\pi}}^\pi$ represents the total spin and parity of the intermediate states. Note that when the orbitals are fully filled, $\delta_{C2}^{p-shell}$ is solely induced by $\Lambda^{\tilde{\pi}}_{k_ak_b}\ne 0$. Therefore, $\mathcal{M}_0$ is unaffected by this radial mismatch-induced core-orbital contribution, 
as it is evaluated with an isospin-symmetry basis. 
The $\mathcal{M}_0$ values obtained in the $sd$ shell will be employed in our evaluation of $\delta_{C2}^{p-shell}$. 

For the present study, we consider only ten intermediate states, specifically the first $\frac{1}{2}^-$ state, the lowest four $\frac{3}{2}^-$ states, the lowest two $\frac{5}{2}^-$ states, the lowest two $\frac{7}{2}^-$ states, and the first $\frac{9}{2}^-$ state. Note that the first $\frac{1}{2}^-$ and the first $\frac{9}{2}^-$ states do not contribute to Eq.\,\eqref{p}, although they should be included in the sum rule\,\eqref{FM1}. The excitation energies of these intermediate states, denoted collectively as $E_x^{\tilde{\pi}}$, to be used for evaluating the radial mismatch factors, are assumed to be identical for the $\beta^-$ and $\beta^+$ decays, and are taken from experimental data for $^{27}$Al\,\cite{NNDCENSDF}. 
Excitation energies for negative-parity states in $^{27}$Si are less known. The spectroscopic factors for higher negative-parity intermediate states are neglected. 
With Woods-Saxon radial wave functions, we found that the radial mismatch factor decreases almost linearly with intermediate-state energy. For the transition $^{28}$P($\beta^+$)$^{28m}$Si, we obtain $\Lambda_{p_\frac{3}{2}p_\frac{3}{2}}^{\tilde{\pi}}=-0.129E_x^{\tilde{\pi}}+1.921\,\%$, $\Lambda_{p_\frac{1}{2}p_\frac{1}{2}}^{\tilde{\pi}}=1.319\Lambda_{p_\frac{3}{2}p_\frac{3}{2}}^{\tilde{\pi}}$, $\Lambda_{p_\frac{1}{2}p_\frac{3}{2}}^{\tilde{\pi}}=0.958\Lambda_{p_\frac{3}{2}p_\frac{3}{2}}^{\tilde{\pi}}$, and $\Lambda_{p_\frac{3}{2}p_\frac{1}{2}}^{\tilde{\pi}}=1.417\Lambda_{p_\frac{3}{2}p_\frac{3}{2}}^{\tilde{\pi}}$. Whereas, for the mirror process of $^{28}$Al($\beta^-$)$^{28m}$Si, we obtain $\Lambda_{p_\frac{3}{2}p_\frac{3}{2}}^{\tilde{\pi}}=-0.031E_x^{\tilde{\pi}}+0.464\,\%$, $\Lambda_{p_\frac{1}{2}p_\frac{1}{2}}^{\tilde{\pi}}=1.220\Lambda_{p_\frac{3}{2}p_\frac{3}{2}}^{\tilde{\pi}}$, $\Lambda_{p_\frac{1}{2}p_\frac{3}{2}}^{\tilde{\pi}}=0.842\Lambda_{p_\frac{3}{2}p_\frac{3}{2}}^{\tilde{\pi}}$, and $\Lambda_{p_\frac{3}{2}p_\frac{1}{2}}^{\tilde{\pi}}=1.599\Lambda_{p_\frac{3}{2}p_\frac{3}{2}}^{\tilde{\pi}}$. The unit of $E_x^{\tilde{\pi}}$ in these expressions is MeV. 
\begin{table}[ht!]
\centering
\caption{$p$-shell contribution to the radial mismatch correction and the corresponding asymmetry. All listed numbers are in \% unit. } 
\label{tab1} 
\begin{tabular*}{\linewidth}{c @{\extracolsep{\fill}}|c|c|c} 
\toprule
Scenarios	& $\delta_{C2}^{p-shell}(\beta^+)$	& $\delta_{C2}^{p-shell}(\beta^-)$	& $\delta^{p-shell}$ \\
\midrule
I	&-7.193	&-1.717	&-5.476 \\
II	&-7.071	&-1.727	&-5.344 \\
III	&-5.371	&-1.313	&-4.058 \\
IV	&-8.23	&-1.945	&-6.285 \\
V	&-5.299	&-1.299	&-4 \\
VI	&-4.846	&-1.138	&-3.708 \\
VII	&-6.248	&-1.535	&-4.713 \\
VIII &-2.467	&-0.586	&-1.881 \\
IX	&-4.791	&-1.143	&-3.648 \\
\midrule
Average	&-3.551$\pm$1.699 &-2.862$\pm$0.408	&-4.346$\pm$1.291 \\
\bottomrule
\end{tabular*}
\end{table}

We explore the individual spectroscopic factors in the sum rules in Eq.\,\eqref{FM1} and \eqref{FM2} under nine different scenarios, without performing additional microscopic calculations. 
\begin{itemize}
    \item \textbf{Scenario I:} Each term  on the left-hand side of both Eq.\eqref{FM1} and \eqref{FM2} holds an equal weight, reflecting constantly distributed spectroscopic factors. 
    \item \textbf{Scenario II:} Both $S^T(f;\tilde{\pi} k_a)$ and $S^T(i;\tilde{\pi} k_b)$ gradually decrease as a function of excitation energy, with each term in the sum rules being half of the previous term. 
    \item \textbf{Scenario III:} The  reverse of the scenario II, wherein both $S^T(f;\tilde{\pi} k_a)$ and $S^T(i;\tilde{\pi} k_b)$ gradually increase as a function of excitation energy, doubling the value of each term in the sum rules compared with the previous one. 
    \item \textbf{Scenario IV:} The spectroscopic factors $S^T(f;\tilde{\pi} k_a)$ behaves as in the scenario I whereas $S^T(i;\tilde{\pi} k_b)$ as in the scenario II. 
    \item \textbf{Scenario V:} The spectroscopic factors $S^T(f;\tilde{\pi} k_a)$ behaves as in the scenario II whereas $S^T(i;\tilde{\pi} k_b)$ as in the scenario I. 
    \item \textbf{Scenario VI:} The spectroscopic factors $S^T(f;\tilde{\pi} k_a)$ behaves as in the scenario I whereas $S^T(i;\tilde{\pi} k_b)$ as in the scenario III. 
    \item \textbf{Scenario VII:} The spectroscopic factors $S^T(f;\tilde{\pi} k_a)$ behaves as in the scenario III whereas $S^T(i;\tilde{\pi} k_b)$ as in the scenario I. 
    \item \textbf{Scenario VIII:} The spectroscopic factors $S^T(f;\tilde{\pi} k_a)$ behaves as in the scenario II whereas $S^T(i;\tilde{\pi} k_b)$ as in the scenario III.
    \item \textbf{Scenario IX:} The spectroscopic factors $S^T(f;\tilde{\pi} k_a)$ behaves as in the scenario III whereas $S^T(i;\tilde{\pi} k_b)$ as in the scenario II.
\end{itemize} 

An increase of spectroscopic factors is also considered, although it might seem unlikely, since this observable generally decreases on average with excitation energy. We expect that these nine scenarios would be sufficiently diverse to cover all the possible variations of spectroscopic factors across the considered intermediate states. 
The results for $\delta_{C2}^{p-shell}$ and the asymmetry contribution for each scenario are presented in Table\,\ref{tab1}. These results vary remarkably across scenarios. Nevertheless, all consistently yield a negative $p$-shell contribution to the asymmetry parameter, leading to an improved agreement with the experimental data. 
Combining the averaged $\delta^{p-shell}$ in Table\,\ref{tab1} with the contributions from the $sd$ shell and from the isospin-mixing correction, the total value for the asymmetry parameter becomes 3.664$\pm$3.245\,\%.
This total asymmetry value is still larger than the experimental value by about one standard deviation. However, it is more than two times smaller than the value obtained without the $p$-shell contribution. 

\section{Conclusion}\label{conclusion}

We demonstrate that the experimentally observed negative asymmetry in Gamow-Teller transition rates between $^{28}$Al($\beta^-$)$^{28m}$Si and $^{28}$P($\beta^+$)$^{28m}$Si, can be reasonably explained by integrating the $p$-shell contribution, estimated using the French-Macfarlane sum rules. We consider various scenarios of individual spectroscopic factor variations across the included intermediate states, covering the range of possibility. We also improve the calculated values of the isospin-mixing correction by scaling them with experimental energy separation between the admixed states. Nevertheless, the isospin-mixing contribution is almost negligible. The present result for the total asymmetry parameter is over 50\,\% lower than the previous $sd$-shell value. 
The remaining discrepancy with experimental data of one standard deviation might be attributed to the absent radiative corrections, or the existence of more exotic distributions of spectroscopic factors. 
An exact first-principle calculations of the $\delta_C$ correction as well as data on the radiative corrections would be extremely useful for a comparison. 
Experimental data regarding the spectroscopic factors for the removal of a nucleon from both initial and final states can serve as a filter to identify intermediate states that make significant contributions, provided such data are accessible. 
The radial mismatch correction for Gamow-Teller transitions has a structural behavior that is potentially sensitive to the contribution from deep-lying orbitals. Our approach represents a powerful tool for estimating core-orbital contribution where the first-principle calculation is not accessible. 

\section*{Acknowledgements}

L.\,Xayavong thanks M.\,Brodeur and N.\,Severijn for a fruitful discussion on the evaluation of $ft$ values. 
L.\,Xayavong and Y.\,Lim are supported by the National Research Foundation of Korea(NRF) grant funded by the Korea government(MSIT)(No. 2021R1A2C2094378). Y.\,Lim is also supported by the Yonsei University Research Fund of 2023-22-0126. 


\end{document}